# Superconductivity in undoped $T'$-$RE_2$CuO$_4$ with $T_c$ over 30 K

O. Matsumoto [a], A. Utsuki [a], A. Tsukada [b], H. Yamamoto [c], T. Manabe [d], *M. Naito [a]

[a] Department of Applied Physics, Tokyo University of Agriculture and Technology

Naka-cho 2-24-16, Koganei, Tokyo 184-8588, Japan

[b] Geballe Laboratory for Advanced Materials, Stanford University,

Stanford, California 94305, USA

[c] NTT Science and Core Technology Laboratory Group, NTT Corporation, 3-1

Morinosato-Wakamiya, Atsugi, Kanagawa 243-0198, Japan

[d] National Institute of Advanced Industrial Science and Technology (AIST)

Higashi 1-1-1, Tsukuba, Ibaraki 305-8565, Japan

**Abstract**

In this article, we report the superconductivity in $T'$-$RE_2CuO_4$ ($RE$ = Pr, Nd, Sm, Eu, and Gd), which have been for a long time believed as a Mott insulator. Our discovery was achieved by using metal-organic decomposition (MOD), an inexpensive and easy-to-implement thin-film process. The keys to prepare the superconducting films are firing with low partial-pressure of oxygen and reduction at low temperatures. The highest $T_c$ of undoped $T'$-$RE_2CuO_4$ is over 30 K, substantially higher than "electron-doped" analogs. Remarkably, $Gd_2CuO_4$, even the derivatives of which have not shown superconductivity so far, gets superconducting with $T_c^{onset}$ as high as 20 K. The implication of our discovery is briefly discussed.

*) Corresponding author. Address: Department of Applied Physics, Tokyo University of Agriculture and Technology, Naka-cho 2-24-16, Koganei, Tokyo 184-8588, Japan. Tel. +81 42 388 7229; fax: +81 42 385 6255. E-mail address: minaito@cc.tuat.ac.jp.

**Introduction**

The lanthanide copper oxides of a general formula $RE_2CuO_4$ ($RE$ = rare-earth element) crystallize in two different structures: the $K_2NiF_4$ ($T$) structure with octahedral $CuO_6$ coordination and the $Nd_2CuO_4$ structure ($T'$) with square-planar $CuO_4$. It has been believed that, in either case, the undoped ($Cu^{2+}$) mother compounds are a Mott insulator, and that superconductivity develops only after $p$-type doping in the $T$ structure or after $n$-type doping in the $T'$ structure. Against this belief, in this article, we demonstrate that superconductivity is achieved in $T'$-$RE_2CuO_4$ without doping. Our discovery contradicts with the past results supporting undoped $T'$-cuprates to be insulating. The clue to understanding the sharp contrast between the past and our results is impurity oxygen ($O_{ap}$) at the apical site. The superconductivity in the $T'$ cuprates is very sensitive to the impurity oxygen, which is known to be a very strong pair breaker as well as a very strong scatterer [1]. Therefore the generic phase diagram of the $T'$-cuprates can be reached only after complete removal of $O_{ap}$ atoms.

The first attempt along this line was made by Brinkmann *et al.* in 1995 [2]. They reported that the superconducting (SC) window for $Pr_{2-x}Ce_xCuO_4$ expands by an improved reduction process down to $x = 0.04$ even with slightly increasing $T_c$. The expanded SC window can be explained by the suppression of $O_{ap}$-induced antiferromagnetic (AF) order. Employing a thin film process, we also demonstrated that the superconducting range in $T'$-$La_{2-x}Ce_xCuO_4$ is extended down to $x = 0.045$ [3], and also that the end-member $T'$-$La_2CuO_4$, which has been supposed to be a Mott insulator, shows metallic resistivity down to 150 K with $\rho(300\ K)$ as low as 2 m$\Omega$cm [4]. Furthermore, in the previous article, we reported the superconductivity with $T_c \sim 25$ K in "undoped" $T'$-$(La,RE)_2CuO_4$, which was achieved by state-of-the-art molecular beam epitaxy (MBE) technique [5]. However, there has been a great controversy on interpretation of the superconducting nature in these cuprates: truly undoped or electron-doped due to oxygen deficiencies. One drawback to obtaining definitive conclusion is that *T'*-(La,*RE*)$_2$CuO$_4$ is difficult to synthesize by bulk processes. Therefore they are not amenable to any measurement of oxygen nonstoichiometry like chemical analysis or neutron diffraction. Hence only speculative discussions have been made based on transport measurements such as Hall coefficient and Nernst signal [6, 7].

In this article, we report the superconductivity with $T_c$ over 30 K in *T'*-*RE*$_2$CuO$_4$ (*RE* = Pr, Nd, Sm, Eu, and Gd). This was achieved by metal-organic decomposition (MOD), a rather inexpensive and easy-to-implement thin film process with no fundamental difference from bulk synthesis, which is in contrast to leading-edge MBE required to produce the first generation of undoped superconductors, *T'*-(La,*RE*)$_2$CuO$_4$. *T'*-*RE*$_2$CuO$_4$ has been believed as a Mott insulator since the discovery of "electron-doped" superconductors, *T'*-(*RE*,Ce)$_2$CuO$_4$ in 1989 [8]. However, our new synthetic route with low-$P_{O2}$ firing and low-temperature reduction has made them superconducting.

**Experimental**

The superconducting *RE*$_2$CuO$_4$ thin films were prepared by the MOD method by using *RE* and Cu naphtenate solutions [9]. The solutions were mixed with the stoichiometric ratio and the mixed solution was spin-coated on substrates. As a substrate, we used SrTiO$_3$ (STO) (100) and DyScO$_3$ (DSO) (110) [10]. The coated films were first calcined at 400 $^o$C in air to obtain precursors. Then the precursors were fired at 850 – 900 $^o$C in a tube furnace under a mixture of O$_2$ and N$_2$, varying the oxygen partial pressure $P_{O2}$ from 2 x 10$^{-4}$ atm to 1 atm. Finally the films were "reduced" in vacuum (<10$^{-4}$ Torr ≈ 10$^{-7}$ atm) at various temperatures for removal of O$_{ap}$. Films with no reduction were also prepared for a reference, and named as "as-grown". After the reduction, the films were furnace-cooled in vacuum to avoid re-absorption of oxygen. The film thickness was typically 800 Å. The crystal structure and the *c*-axis lattice parameter ($c_0$) of the films were determined with a powder X-ray diffractometer, and the resistivity was measured by a standard four-probe method.

**Results**

Figure 1 shows the X-ray diffraction (XRD) patterns of typical films prepared by MOD. All peaks are sharp and can be indexed to (00$l$) of the $Nd_2CuO_4$ structure, indicating that the films are single-phase $T'$, and also single-crystalline as achieved via solid-state epitaxy. The (008) peaks shift systematically toward larger $2\theta$ from Pr to Gd, indicating that the $c$-axis shortens in this order.

As well known, the end-member $T'$ cuprates are semiconducting with standard bulk synthesis. In contrast, they become fairly metallic in thin film form, especially for large $RE^{3+}$ ions like La, Pr, Nd and Sm. This is, we believe, because thin films may be advantageous due to a large surface-to-volume ratio in removing apical oxygen atoms with regular oxygen atoms intact. However, our many-year attempts to obtain superconductivity by optimizing post-reduction conditions have been unsuccessful except for the MBE-grown $T'$-$(La,RE)_2CuO_4$. Then, in this study, we attempted another recipe of growing films in a low-$P_{O2}$ atmosphere, the aim of which is to minimize the amount of impurity oxygen in advance before the post-reduction process.

Figure 2 shows the effect of $P_{O2}$ during firing, which compares the resistivity of two $Sm_2CuO_4$ films (A and B): film A fired at 900°C for 1 hour in $P_{O2}$ = 1 atm, followed by reduction at 750°C in 10 min, and film B fired at 850°C for 1 hour in $P_{O2}$ = 2.8 x $10^{-3}$atm, followed by reduction at 440°C in 10 min. Film A is metallic down to 180 K with $\rho$(300 K) ~ 100 mΩcm, but shows upturn at lower temperatures. By contrast, film B is all the way metallic with $\rho$(300 K) as low as 900 μΩcm, and shows superconductivity at $T_c^{onset}$ = 28 K ($T_c^{end}$ = 25K). In both the films, the reduction temperature and time were optimized, which means there is a substantial difference in the optimal reduction temperature. Films fired in low-$P_{O2}$ do not stand against reduction at temperatures higher than 500 °C whereas films fired in $P_{O2}$ = 1 atm do against reduction temperatures as high as 750 °C. Namely the phase stability line appears to shift toward lower temperature or equivalently higher $P_{O2}$ with decreasing the amount of $O_{ap}$ atoms.

The dramatic effect of low-$P_{O2}$ synthesis is actually corroborated from the structural aspect. Figure 3 demonstrates evolution of the $c_0$ of two series of $Sm_2CuO_4$ films as a function of reduction time. One series of the films were prepared in $P_{O2} = 1$ atm, followed by reduction at 750°C. The others were prepared in $P_{O2} = 2.8 \times 10^{-3}$ atm, followed by reduction at 440°C. The $c_0$ for the as-grown film (plotted at $t_{red} = 0$) prepared in $P_{O2} = 1$ atm shows a good agreement with the previously reported bulk value [11] whereas the $c_0$ for the as-grown film prepared in $P_{O2} = 2.8 \times 10^{-3}$ atm is significantly (~0.05 Å) smaller. In both the series, the $c_0$ becomes smaller slightly with reduction time. But the initial difference predominates. Taking account of the well-established trend that the $c_0$ increases with the amount of $O_{ap}$ atoms [12], our experimental results indicate that low-$P_{O2}$ synthesis significantly reduces the amount of $O_{ap}$ atoms.

Figure 4 shows the reduction time dependence of the film properties for $Sm_2CuO_4$ films. The data were taken from the films prepared with identical conditions but reduced with a different duration ($t_{red}$). The reduction temperature is fixed at 440°C. The "as-grown" film is semiconducting ($d\rho/dT < 0$) and does not show a superconducting transition. The reduction with $t_{red} = 5$ min is sufficient for the film to be superconducting ($T_c^{onset} \sim 27$ K). The reduction with $t_{red} = 10$ min provides optimal properties, namely highest $T_c$ ($T_c^{onset} = 28$ K, $T_c^{end} = 25$ K) and lowest resistivity [$\rho(300\ K) \sim 900\ \mu\Omega cm$ and $\rho(30\ K) \sim 100\ \mu\Omega cm$]. Further reduction degrades the film properties. The reduction with $t_{red} = 60$ min kills superconductivity and restores semiconducting behavior although there is almost no trace of decomposition as judged from the XRD pattern. Prolonged reduction at slightly higher temperature eventually makes films transparent with the $T'$ structure preserved [13]. It is most likely that excessive reduction leads to loss of oxygen atoms in the $CuO_2$ planes, as supported from the oxygen nonstoichiometry experiments [14]. The results in Fig.

4 indicate that $O_{ap}$ removal proceeds together with loss of oxygen (O1) in the $CuO_2$ planes, but that the former process is slightly quicker than the latter. Hence the reduction with $t_{red}$ = 5 to 30 min yields the situation with $O_{ap}$ substantially removed and still with negligible O1 loss, which achieves superconductivity.

Figure 5(a) shows the superconducting transitions for different *RE*. All *RE*'s we have tested show superconductivity except for *RE* = La, which does not form the *T'* structure by MOD [15]. It should be noted that undoped $T_c$ is substantially higher than electron-doped $T_c$. This is most remarkably demonstrated in $Gd_2CuO_4$: there has been no report that electron-doped $Gd_{2-x}Ce_xCuO_4$ gets superconducting, whereas undoped $Gd_2CuO_4$ has $T_c^{onset}$ as high as 20 K. Figure 5(b) shows a summary of the *RE*-dependence. Although the $T_c$ of electron-doped *T'* shows a steep dependence on *RE*, the $T_c^{onset}$ of undoped *T'* is rather flat, and about 30 K except for Gd. Lower $T_c$ in *RE* = Gd is not intrinsic and most likely due to the material problem that $O_{ap}$ removal is difficult for smaller $RE^{3+}$. We are aware that currently obtained $T_c$ of undoped *T'* is substantially process-dependent, and are not yet sure how much Kelvin the *intrinsic* $T_c$ reaches.

**Discussion**

The key question is the nature of superconductivity in the *nominally undoped* *T'*-$RE_2CuO_4$. Are these materials (a) hole-doped or (b) electron-doped or (c) truly undoped? At first, it is safe to exclude hole-doped superconductivity because it is well-known that excess oxygen makes *T'* parent compounds more insulating, which is in contrast to *T*-$La_2CuO_4$ [4]. Electron doping due to *oxygen deficiencies* is what one can imagine easily, but actually it is unlikely. The oxygen (O1) deficiency in the $CuO_2$ planes may occur, but the golden rule tells us that perfect $CuO_2$ planes are required for achieving high-$T_c$ superconductivity. The oxygen (O2) deficiency in the fluorite $RE_2O_2$ planes is unlikely to occur since solid-state chemistry tells us that the *RE*-O bond is much stronger than the Cu-O bond. Hence, at present, we are thinking that the most likely is undoped superconductivity. This should be established in future works. Since MOD is not fundamentally different from bulk synthesis, superconducting $T'$-$RE_2CuO_4$ may be obtainable in bulk form, which enables oxygen nonstoichiometry measurements such as chemical analysis, neutron diffraction and so on. The answer for undoped or doped can be gained from the probes to the electronic state such as X-ray absorption spectroscopy, photoemission spectroscopy, transport experiments, *etc*.

Finally we explain why the superconductivity in Ce-doped compounds was discovered 20 years ago but that in Ce-free compounds has been discovered just now. Here we describe the role of Ce in previous "electron-doped" superconductors. It is worth to note that Ce doping has a chemical role other than a physical role. Ce in $T'$ cuprates is nearly tetravalent (~+3.8) whereas other $RE$ is trivalent. Hence tetravalent Ce takes a firmer grip of oxygen than trivalent $RE$. This causes two effects. One is a good effect, namely the substantial increase of the binding energy of loosely bound O1 in the $CuO_2$ planes. Accordingly the phase stability lines also shift toward lower $P_{O2}$ or equivalently higher temperature as was demonstrated by Idemoto *et al*. [16]. The other is a bad effect, the increase of the binding energy of harmful $O_{ap}$. If $T'$ cuprates are synthesized in the atmosphere of the Earth, a plenty amount of apical oxygen atoms are introduced in the lattice during firing. In this case, strong reduction is required to clean up $O_{ap}$ atoms, and Ce-doping is good in preventing O1 loss. On the other hand, if $T'$ cuprates are synthesized in low $P_{O2}$, not many apical oxygen atoms are introduced in the lattice during firing. In this case, weak reduction is sufficient, and non-doping is better in that the binding energy of $O_{ap}$ is small. Hence if the synthesis is made in the atmosphere of the Earth, one reaches superconductivity more easily in Ce-doped compounds. But the atmosphere of the Earth is no at all standard in the synthesis of inorganic compounds including oxides. If the synthesis *had been* made in low $P_{O2}$ or other planets, one *could* have reached undoped superconductivity earlier, and the current understanding for the high-$T_c$ superconductivity *should* have been dramatically different. The complex oxygen chemistry in *T'* cuprates has bewildered us off the road toward true high-$T_c$ physics.

**Summary**

We discovered superconductivity in *T'*-$RE_2CuO_4$ (*RE* = Pr, Nd, Sm, Eu, and Gd), which have been believed as a Mott insulator. The synthesis is rather simple and inexpensive, namely low-$P_{O2}$ firing and subsequent low-temperature reduction. One point to be emphasized is that low-$P_{O2}$ phase field has been almost unexplored in the search for new superconductors because of the belief that high $P_{O2}$ *should* be required in the synthesis of $Cu^{2+}$ compounds. The highest $T_c$ of undoped *T'*-$RE_2CuO_4$ is over 30 K, substantially higher than "electron-doped" analogs. It is the most likely that these superconductors are truly undoped although this has to be established in future works.

**Acknowledgements**

The authors thank Dr. Y. Krockenberger and Dr. J. Shimoyama for stimulating discussions, and Dr. T. Kumagai for support and encouragement. They also thank Crystec GmbH, Germany for developing new $RE$ScO$_3$ substrates. The work was supported by KAKENHI B (18340098) from Japan Society for the Promotion of Science (JSPS).

**Figure captions**

Figure 1.    XRD pattern of $RE_2CuO_4$ films prepared by MOD. All peaks can be indexed to the (00$l$) reflections of the $Nd_2CuO_4$ structure, indicating that the films are single-phase $T'$ and also single-crystalline.

Figure 2.    Temperature dependences of resistivity for $Sm_2CuO_4$ films (A and B). Film A was fired at 900 °C for 1 h in $P_{O2}$ = 1 atm followed by reduction at 750°C for 10 min whereas Film B was fired at 850 °C for 1 h in $P_{O2}$ = 2.8 x $10^{-3}$ atm followed by reduction at 440°C for 10 min.

Figure 3.    Firing and reduction condition dependences of $c_0$ for $Sm_2CuO_4$ films. Square symbols are the data from films fired in $P_{O2}$ =1 atm followed by reduction at 750°C and triangle symbols are from films fired in $P_{O2}$ = 2.8 x $10^{-3}$ atm followed by reduction at 440°C. The horizontal axis is reduction time. The dotted line shows the reported bulk value after ref.11.

Figure 4.    Resistivity of $Sm_2CuO_4$ films with different reduction time ($t_{red}$ = 5, 10, 20, 40 and 60min). The films were fired in $P_{O2}$ = 2.8 x $10^{-3}$ atm, and the reduction temperature was fixed at 440°C. The "as-grown" sample was obtained by quenching to ambient temperature in air after firing.

Figure 5.    (a) Resistivity of $T'$-$RE_2CuO_4$ ($RE$ = Pr, Nd, Sm, Eu and Gd) films and (b) ionic radius dependence of $T_c$ ($T_c^{onset}$: ○, $T_c^{end}$: ●).

Fig. 1 Matsumoto et al.

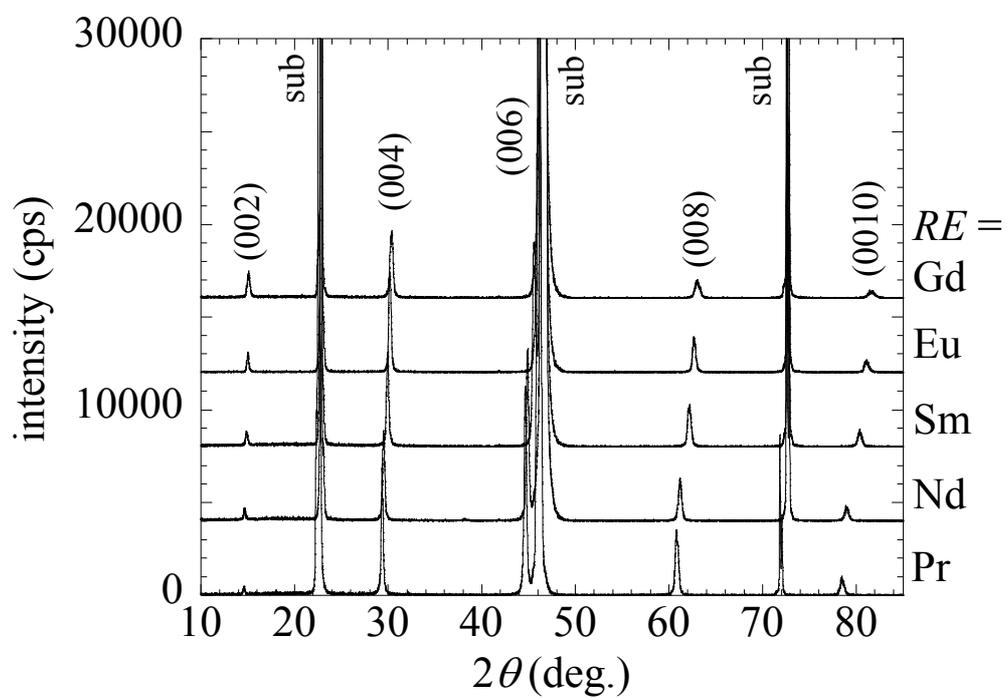

Fig. 2 Matsumoto et al.

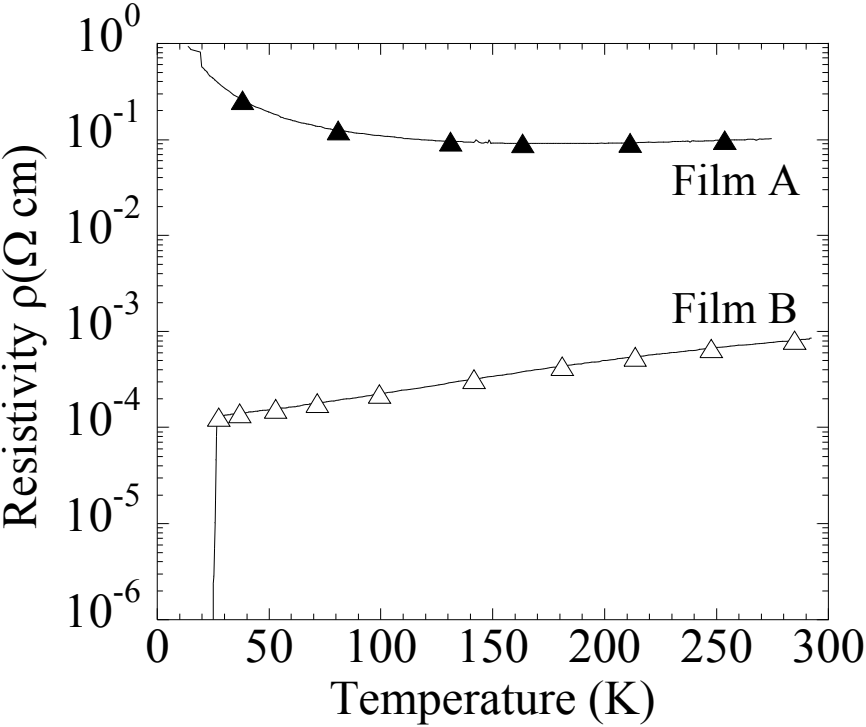

Fig. 3 Matsumoto et al.

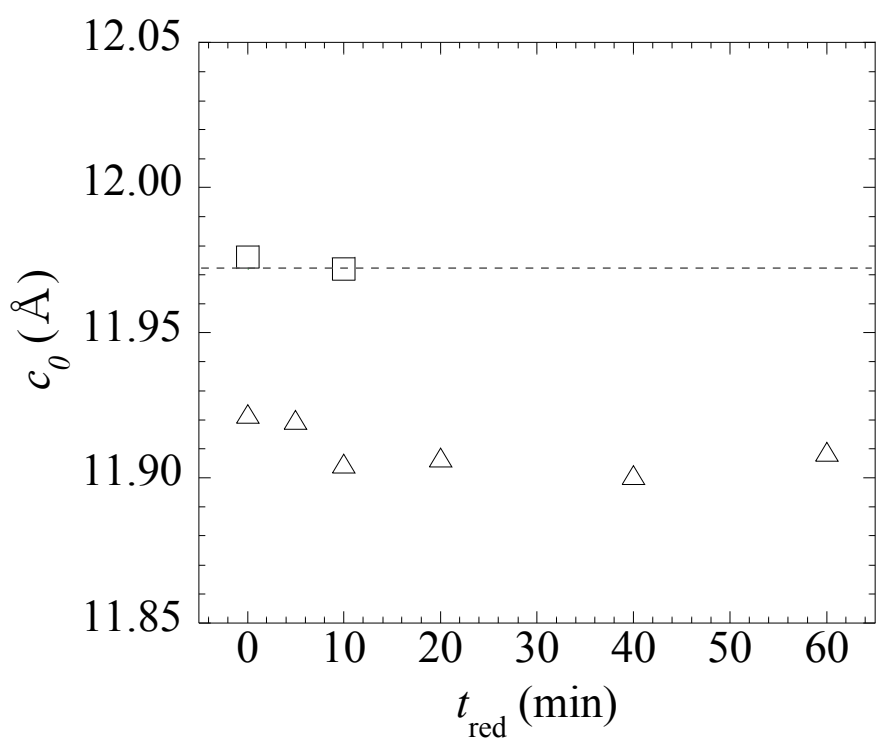

Fig. 4 Matsumoto et al.

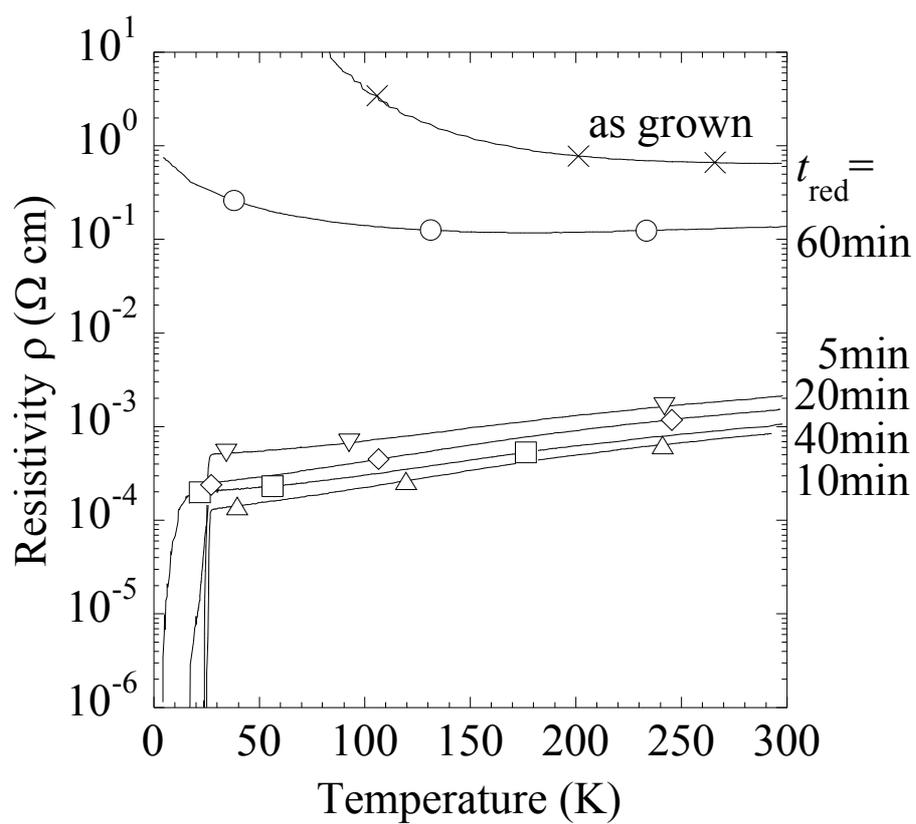

Fig 5 Matsumoto et al.

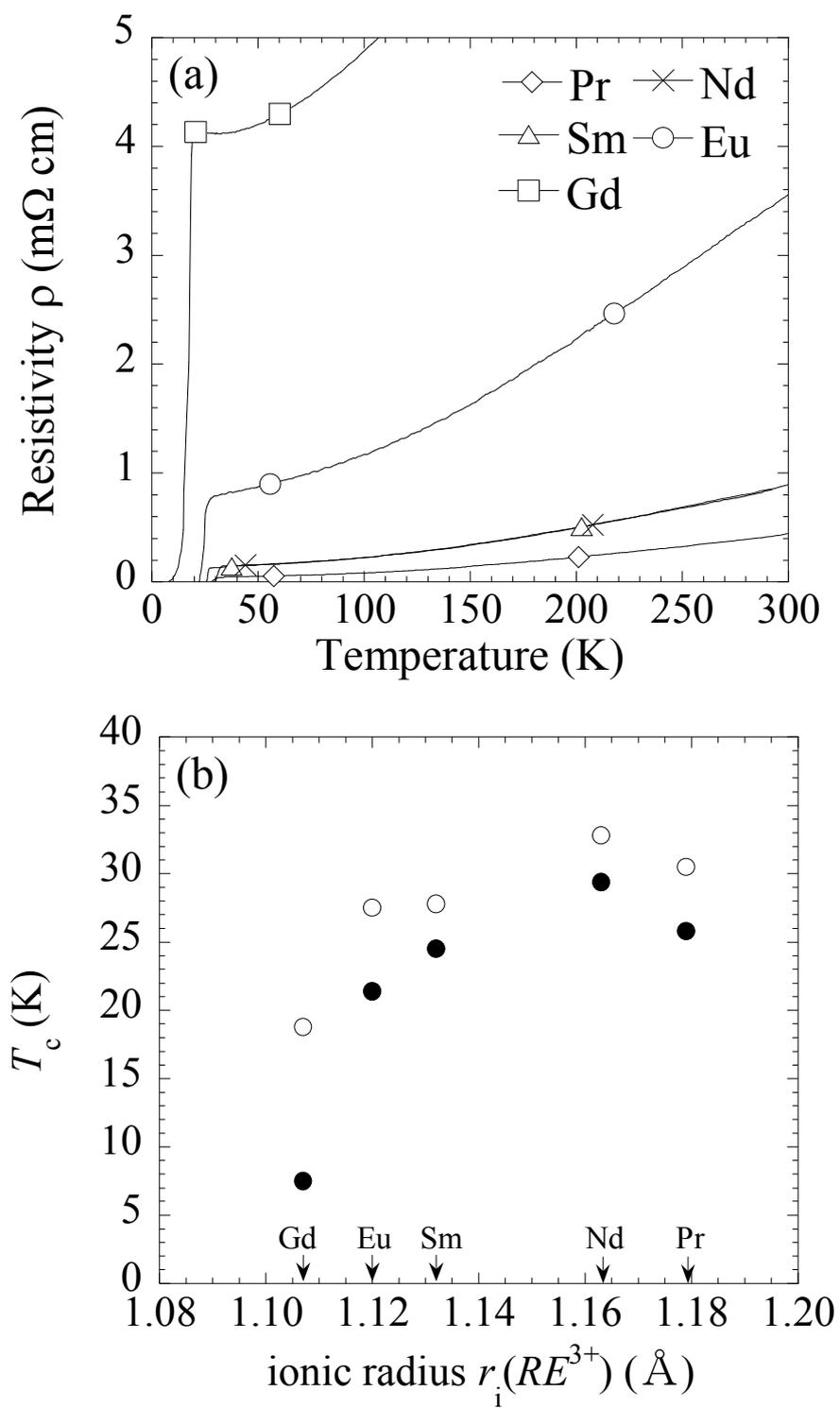